# Field-free platform for topological zero-energy mode in superconductors LiFeAs and PbTaSe$_2$

Songtian S. Zhang[1†], Jia-Xin Yin[1†*], Guangyang Dai[2], Lingxiao Zhao[2], Tay-Rong Chang[3], Nana Shumiya[1], Kun Jiang[4], Hao Zheng[5], Guang Bian[6], Daniel Multer[1], Maksim Litskevich[1], Guoqing Chang[1], Ilya Belopolski[1], Tyler A. Cochran[1], Xianxin Wu[2], Desheng Wu[2], Jianlin Luo[2], Genfu Chen[2], Hsin Lin[7], Fang-Cheng Chou[8], Xiancheng Wang[2], Changqing Jin[2], Raman Sankar[7,8], Ziqiang Wang[4], M. Zahid Hasan[1,9*]

**Affiliations**
[1]Laboratory for Topological Quantum Matter and Advanced Spectroscopy, Department of Physics, Princeton University, Princeton, NJ, USA.
[2]Beijing National Laboratory for Condensed Matter Physics, Institute of Physics, Chinese Academy of Sciences, Beijing, China.
[3]Department of Physics, National Cheng Kung University, Tainan 701, Taiwan.
[4]Department of Physics, Boston College, Chestnut Hill, MA, USA.
[5]Department of Physics and Astronomy, Shanghai Jiaotong University, Shanghai, China
[6]Department of Physics and Astronomy, University of Missouri, Columbia, MO, USA.
[7]Institute of Physics, Academia Sinica, Taipei, Taiwan.
[8]Center for Condensed Matter Science, National Taiwan University, Taipei, Taiwan
[9]Lawrence Berkeley National Laboratory, Berkeley, CA, USA.
[†]These authors contributed equally.
*Corresponding authors. e-mail: jiaxiny@princeton.edu; mzhasan@princeton.edu

**Superconducting materials exhibiting topological properties are emerging as an exciting platform to realize fundamentally new excitations from topological quantum states of matter. In this letter, we explore the possibility of a field-free platform for generating Majorana zero energy excitations by depositing magnetic Fe impurities on the surface of candidate topological superconductors, LiFeAs and PbTaSe$_2$. We use scanning tunneling microscopy to probe localized states induced at the Fe adatoms on the atomic scale and at sub-Kelvin temperatures. We find that each Fe adatom generates a striking zero-energy bound state inside the superconducting gap, which do not split in magnetic fields up to 8T, underlining a nontrivial topological origin. Our findings point to magnetic Fe adatoms evaporated on bulk superconductors with topological surface states as a new platform for exploring Majorana zero modes and quantum information science under field-free conditions.**

The intersection of superconductivity and band topology is emerging as a fertile field of condensed matter and materials research, motivated in part by the search for exotic quantum states and excitations including topological superconductors, helical Dirac fermion Cooper pairing, and the Majorana zero mode (MZM) as well as chiral Majorana edge states [1-14]. Bulk superconductors with topological surface states (TSS) serve as the simplest platforms, as the Fu-Kane hybrid structure [7] originally proposed for generating MZMs in the vortices of an s-wave superconductor proximity coupled to the TSS of a 3D strong TI can be realized in a single material. However, related experimental visualizations of the MZM in the vortex core have been challenging. While by applying an external magnetic field, a zero-energy bound state (ZBS) can be observed in vortices, the existence and stability of the MZM are not guaranteed due to the existence of the very low-energy Caroli-de-Gennes-Matricon vortex core states and vortex mobility. Furthermore, field-induced vortices are difficult to move individually which limits the possibility of manipulating the MZMs for critical operations such as braiding [1-4]. Alternatively, it is known that, without an external field, magnetic impurities can generate in-gap states in superconductors [15, 16] and adatom impurities can be potentially manipulated by the STM tip [17, 18]. Intriguingly, STM measurements observed robust ZBSs on the growth-induced excess Fe atoms in iron-based superconductor Fe(Te,Se) without an applied magnetic field [19]. Subsequent spin-resolved photoemission measurements showed Fe(Te,Se) to host superconducting topological surface states [20], making it reasonable to consider a Majorana interpretation of the spectroscopically observed ZBS [19-21]. Despite these, the superconducting state in Fe(Te,Se) is highly inhomogeneous due to Te/Se alloying [22, 23] and a complex annealing process is required [19]. A more controllable platform with both tunable surface magnetic impurities hosting ZBS and homogeneous Cooper pairing is therefore necessary. In this work, we demonstrate magnetic Fe adatoms deposited on the surface of stoichiometric bulk superconductors with TSS provides the desirable zero-field platform for producing candidate Majorana ZBSs.

Stoichiometric materials LiFeAs and PbTaSe$_2$ have been experimentally shown to be homogeneous superconductors with topological surface states (TSS) [24-30] below $T_C$=17.5K (Fig. 1(a)) and $T_C$=3.8K (Fig. 1(b)), respectively. Theoretical calculations indicate they both posses a non-zero $Z_2$ topological invariant in the band structure, featuring a spin-helical surface state near the Fermi level as shown in the inset of Figs. 1(a) and (b). We are thus motivated to introduce adatom Fe impurities on their surface and explore their atomic scale effects on the superconducting topological surface states. By controlling the deposition rate, we can obtain various adatom Fe concentrations ranging from 0 to 2%, where the atomic nature of the deposited Fe still holds. Figures 1(c) and (d) show the typical topographic images of Fe deposition on the Li surface in LiFeAs and Pb surface in PbTaSe$_2$, respectively, revealing randomly distributed atomic Fe adatoms.

We cool the system to 0.4K to study their local effects on the superconducting ground state. Based on previous studies, LiFeAs is a strong coupling iron-based superconductor with two fully opened energy gaps and slight asymmetry in the low temperature tunneling spectrum [25, 26, 31-33]. On Fe-LiFeAs, we observe a sharp and reproducible zero-energy state at the Fe adatoms (Fig. 2(a), (b)), together with a local suppression of the superconducting coherence peaks, in the absence of external magnetic field. This state is bound to the Fe adatom on the atomic scale as shown in the corresponding zero-energy map in Fig. 2(a) inset and can thus be termed as a ZBS. The ZBS decays rapidly when measuring away from the Fe adatom, but without detectable splitting in energy (Fig.

2(c)). Its spatial decay can be described by an exponential function with a characteristic decay length ~4.7Å, while the ZBS is clearly visible in the raw data within ~16Å around the Fe adatom (Fig. 2(d)). Measuring with high energy resolution, we characterize its full width at half maximum to be approximately 0.44meV (Fig. 2(e)), only 4% of the large superconducting energy gap in LiFeAs ($2\Delta$=11.6meV). We further subject the ZBS to external vector magnetic fields and find that both the ZBS and the superconducting coherent peaks weaken (Fig. 2(f)) without any detectable energy-shift or splitting of the ZBS. These systematic experimental results support that isolated Fe adatoms on LiFeAs surface generate robust non-splitting ZBSs.

To resolve the excitations within the much smaller energy gap in PbTaSe$_2$, we use superconducting tips prepared *in situ* by contacting a Pt/Ir tip with the Pb surface of this material. A superconducting tip is advantageous, especially in the case of a small-gap superconductor, as it enhances the density of states signal from the coherent peaks and in-gap states [14, 34-36]. Here, the coherence peaks of the sample's superconducting gap $\Delta_S$ occur at energy $\pm(\Delta_T + \Delta_S)$ where $\Delta_T$ is the superconducting gap of the tip. An in gap state of the sample at an energy ε would occur at $\pm(\Delta_T + \varepsilon)$, because an unpaired electron remains in the tip, giving an offset to the sample resonances by $\Delta_T$ [14, 34].

The tunneling conductance spectrum obtained on the pristine surface as shown in Fig. 3(a) shows a sharp fully gapped structure. It is immediately evident that the coherence peaks are both much stronger and at larger energies (±1.2meV) compared to spectra measured with a normal tip (±0.5meV). Both features are consistent with tunneling from a superconducting tip possessing an energy gap $\Delta_T$ of 0.7meV [14, 34-36], into a superconductor with a pairing energy gap $\Delta_S$=0.5meV. To further support this identification, we apply an external magnetic field along the c-axis. Interestingly, this rapidly reduces the gap size to 0.7meV for a field of 0.1T and saturates at this value for spectra taken far from any vortices (Fig. 3(a)). Further increasing the field, the gap gradually fills in and vanishes entirely at B=2T (Fig. 3(a) and its inset), noticeably higher than the $H_{C2}$~0.087T [30] for PbTaSe$_2$. These observations confirm the 0.7meV gap originates from the superconducting tip, while the 0.5meV gap originates from the superconducting state of the sample, which can be destroyed by a much weaker magnetic field of 0.1T, comparable to the upper critical field $H_{C2}$ of PbTaSe$_2$. The persistence of the tip gap to a much higher field is consistent with previous experiments [37]. The zero-energy vortex core state [29] provides another assessment of the tip gap. We measure the vortex core states generated under a field-cooled technique at B=0.04T (Fig. 3(b) inset). Comparing the dI/dV spectra taken on and off a vortex core as shown in Fig. 3(b) clearly reveals the presence of electronic states at ±0.7meV=±$\Delta_T$, corresponding to "zero-energy" vortex core states around ε=0.

We next deposit dilute Fe impurities on the surface and measure the spectra using the same superconducting tip. Measuring the spectrum at a Fe adatom (Fig. 3(c) inset), a pronounced in-gap state emerges at ±0.7meV=±$\Delta_T$ with a strong concomitant suppression of the coherence peaks (Fig. 3(c)). A subtraction of the two curves shows the spectral weight transfer from the coherence peak to the in-gap states (Fig. 3(d)). This in-gap state is also an intrinsic ZBS associated with a defect excitation in the superconducting state. A spatial measurement of the spectra across this Fe impurity (Fig. 3(e)) demonstrates the rapid decay of the in-gap state on the atomic scale without discernable spitting or dispersion. A more detailed analysis in Fig. 3(f) shows this state is tightly bound spatially to the Fe adatom within ~10Å, together with a local suppression of the coherence peaks. Crucially, we find every isolated Fe adatom examined exhibits an in-gap state at the same energy $\Delta_T$, equivalent to the "zero-energy" $\varepsilon=0$ in the superconducting state of $PbTaSe_2$ (Fig. 3(g)), within our energy resolution (Fig. 3(g) inset), essentially demonstrating that each Fe adatom robustly generates a non-splitting ZBS. To further elucidate these remarkable properties, we numerically deconvolute the tunneling spectra, supporting that the Fe adatom generates a sharp ZBS (Fig. 3(h)).

We now discuss the nature of the ZBS induced by the Fe adatom in both LiFeAs and $PbTaSe_2$. In a conventional s-wave superconductor, the magnetic impurity generates Yu-Shiba-Rusinov (YSR) states [15, 16]. The number of YSR states depends on the atomic orbital nature of individual magnetic impurities [38], and the Fe impurity often leads to multiple in-gap YSR states in STM experiments [39-41]. Moreover, one set of YSR states generally features two bound states with equal energies away from zero-energy. Under a magnetic field, the YSR state will show a Zeeman splitting which in this case should be on the order of 1meV for an external 8T field, which we do not see [40]. Even for a quantum impurity the resonance peak is usually off zero-energy due to the existence of the finite potential scattering and should also exhibit a magnetic-field induced Zeeman-like splitting [16].

A recent theoretical proposal [21] offers a heuristic understanding of the observed phenomenon, arguing that magnetic ions in a s-wave superconductor with strong spin-orbit coupling and topological surface states may generate vortex-like objects without external magnetic fields and host robust MZMs localized at the magnetic ion sites. In this theory, the phase winding of the Cooper pairs develops spontaneously around the magnetic impurity, and the role of the magnetic field is played by the spin-orbit exchange field. The MZMs arise robustly in the spontaneous vortex as a sharp ZBS in the STM conductance spectra, since the exchange field pushes the nonzero angular momentum low-energy vortex core states to the SC gap edges, reducing the quasiparticle contamination and stabilizing the MZM, consistent with the absence of ZBSs in magnetic field-induced vortices in LiFeAs [31-33] and the appearance of sharp ZBSs at the Fe adatoms in the same material. Furthermore, even when the Dirac point of the topological surface states is located far from the Fermi energy as in $PbTaSe_2$, the exchange interaction can still push the in-gap states away from zero energy, leading to the much sharper zero-bias conductance peak observed at the adatom Fe site compared to the field induced vortex core. Our current observations suggest the key criteria for this theory including the magnetic ions, s-wave like superconductivity, spin-orbit coupling, topological surface state and ZBS can be satisfied in both adatom Fe-LiFeAs and Fe-$PbTaSe_2$ systems, lending credence to a topological interpretation of the observed zero-energy states. Crucially, these new systems hold a great advantage over Fe(Te,Se) in the simplicity with which we can, without affecting the bulk homogeneous superconductivity, generate these magnetic impurities on the surface of this stoichiometric crystal. Through these studies of different superconductors with and without inversion symmetry, in the weak and the strong coupling limit, and for interstitial and adatom Fe, we unveiled the rich and robust phenomena of magnetic Fe impurity induced zero-energy excitations as candidate Majorana zero modes on bulk superconductors with a non-zero $Z_2$ topological invariant in the band structure.

**Figures and Tables**

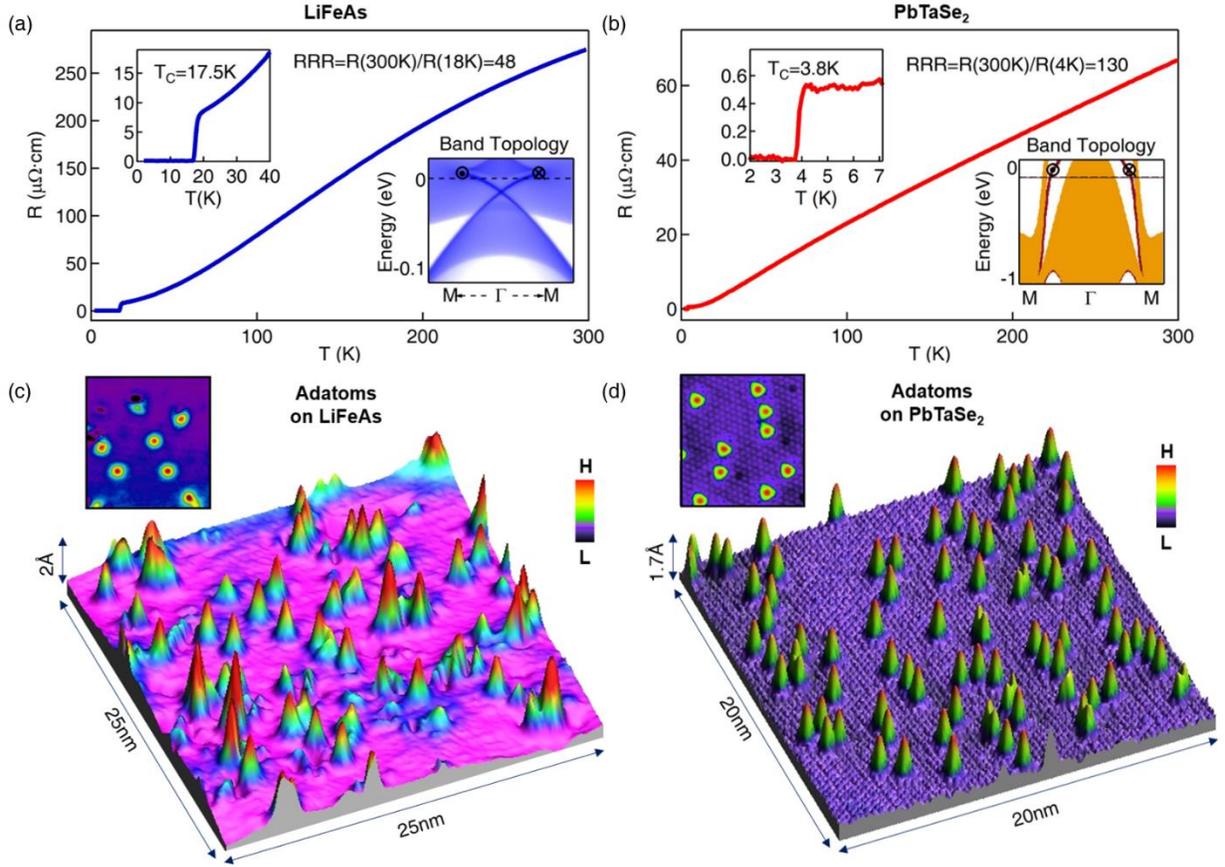

**Fig. 1.** (a), (b) Transport measurement of LiFeAs and PbTaSe$_2$ respectively, showing a large residual-resistance ratio of 48 and 130, respectively, indicating the high quality of the samples. Inset left: enlarged resistivity curve near the superconducting transition temperature. Inset right: calculated band structure along a high symmetry direction, with the topological surface state featuring spin-momentum locking. (c), (d) Topographic 3D image of randomly scattered Fe adatoms on the LiFeAs and PbTaSe$_2$ surface respectively, demonstrating the atomic nature of the Fe deposition at cryogenic conditions. Insets: 2D zoom-in view of adatoms.

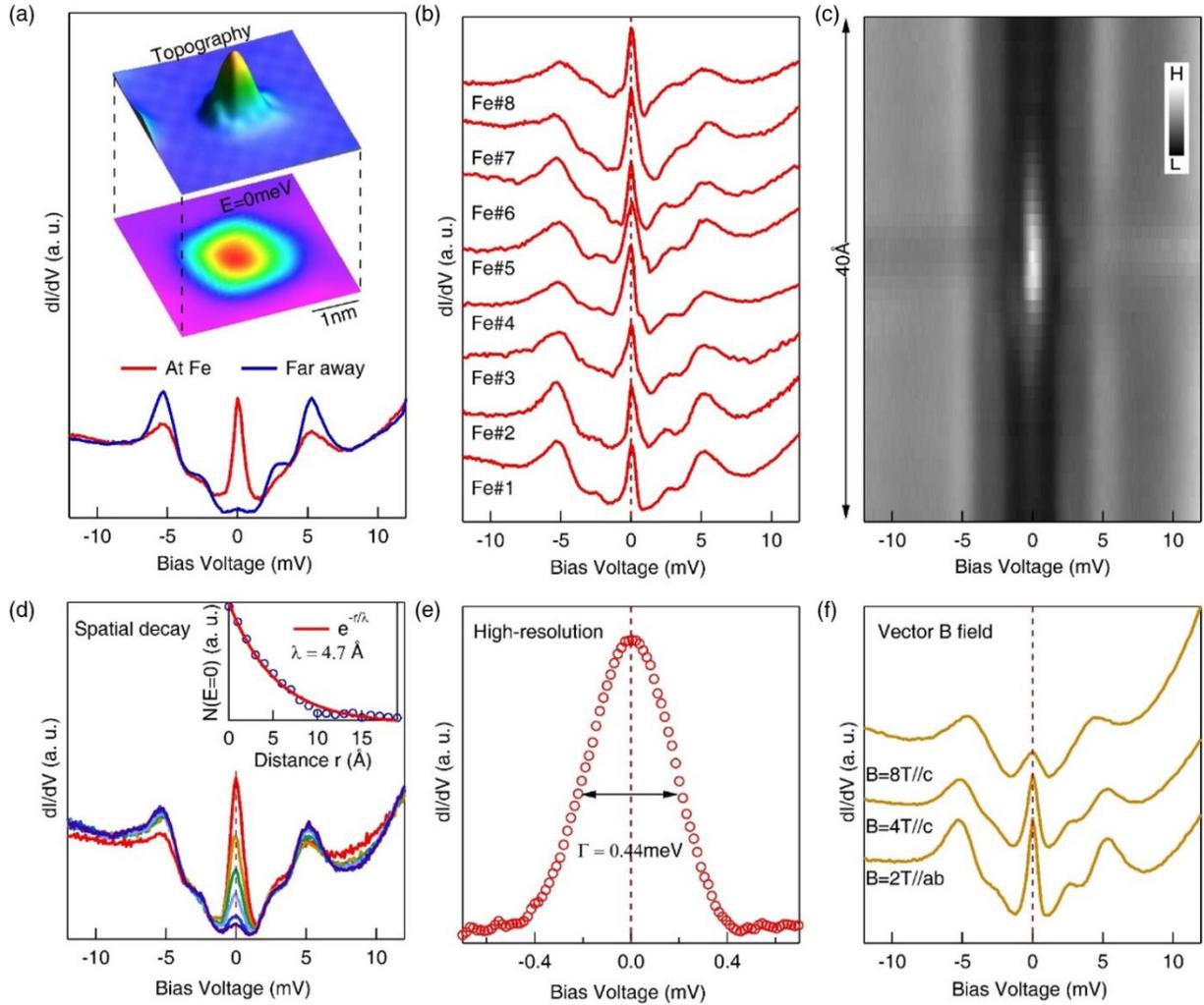

**Fig. 2.** (a) Impurity state of a Fe adatom, featuring a zero-energy state. Inset: topographic image of the impurity (upper) and its dI/dV map at zero-energy (lower). (b) dI/dV spectra on eight different Fe adatoms. (c) Intensity map of dI/dV spectrum across the Fe adatom, showing decay of the zero-energy state without detectable splitting. (d) Selected spectrums from (c). Inset: spatial evolution of the zero-energy state. (e) High energy resolution dI/dV spectrum of the zero-energy state. (f) Vector magnetic field perturbation to the zero-energy state, showing suppression of the coherence of the spectrum.

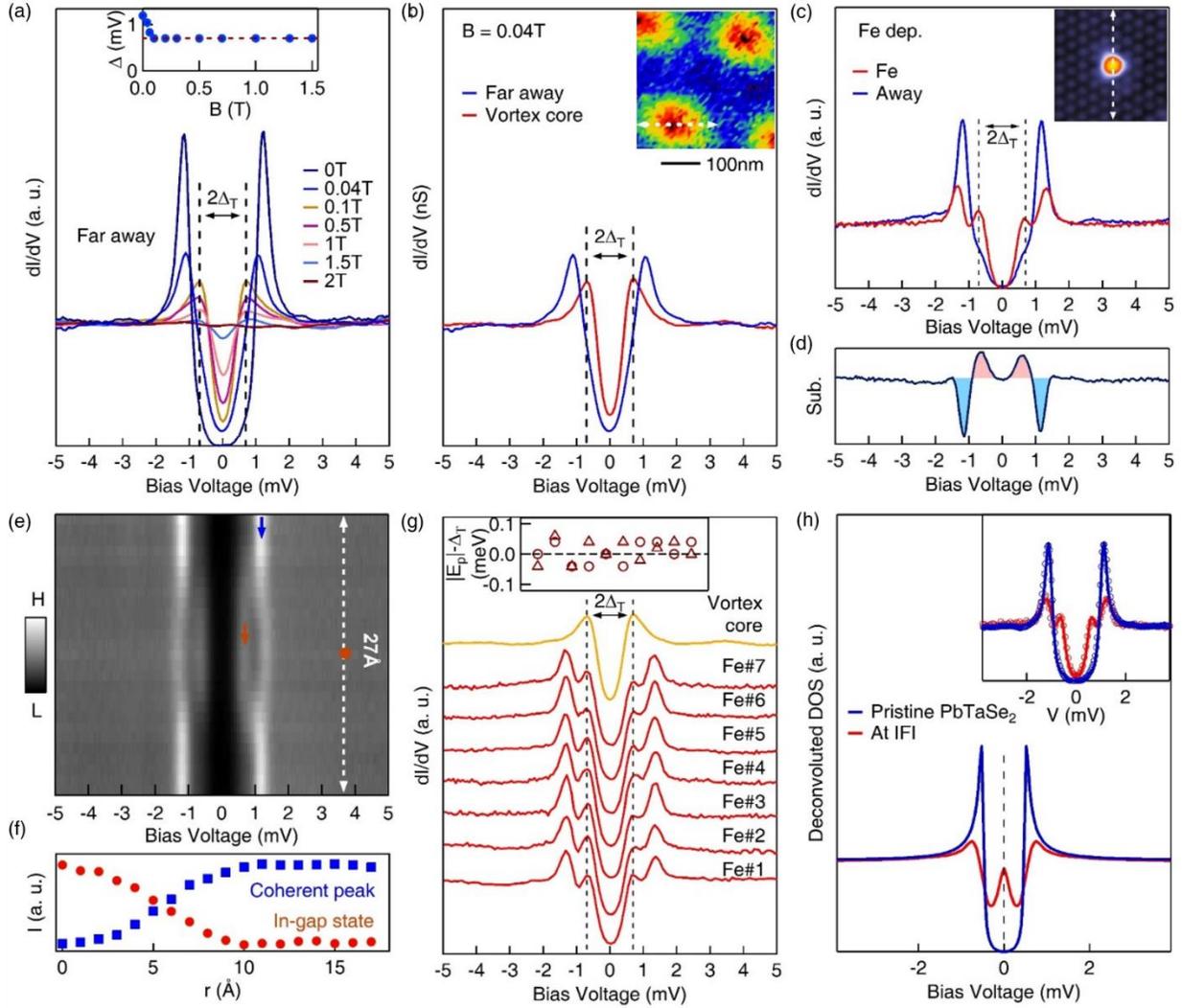

**Fig. 3.** (a) dI/dV spectra taken far from any vortices on a pristine sample as a function of magnetic field. Inset: gap evolution with field. (b) dI/dV spectra on and far from a vortex core (imaged inset) at B=0.04T. Inset: dI/dV map at $\Delta_T$=0.7mV. (c) dI/dV spectra at a Fe adatom (red) and away (blue). Inset: topographic image of Fe adatom. (d) Subtraction of the two spectra in (c), characterizing the emergence of a pair of in-gap states (red) at $\pm\Delta_T$ ($\varepsilon$=0). (e) Intensity map of dI/dV spectrum taken across the Fe adatom. (f) Spatial evolution of the coherence peak and in-gap state intensity. (g) dI/dV spectra on seven Fe adatoms (red), showing the robust in-gap state at $\pm\Delta_T$ ($\varepsilon$=0); spectrum on vortex core (orange) exhibits a peak at the same energy. Inset: deviation from zero-energy for ten measured IFIs (offset for clarity). Circles (triangles) denote bound states at the negative (positive) energy. (h) Deconvoluted tunneling spectra of pristine PbTaSe$_2$ (blue) and at the Fe adatom (red) with both superconducting tip gap and thermal broadening effect at 0.4K subtracted. Inset: spectra convoluted with a superconducting tip gap and Fermi distribution function at 0.4K, plotted as open blue and red circles, respectively.

**Acknowledgments:**

**Funding:** Experimental and theoretical work at Princeton University was supported by the Gordon and Betty Moore Foundation (GBMF4547/ Hasan) and the United States Department of Energy (US DOE) under the Basic Energy Sciences program (grant number DOE/BES DE-FG-02-05ER46200). Z.W and K.J. acknowledge US DOE grant DE-FG02-99ER45747. G.B. is supported by the U.S. National Science Foundation (NSF-DMR-0054904). T.-R.C. acknowledges MOST Young Scholar Fellowship (MOST Grant for the Columbus Program number 107-2636-M-006-004-), the National Cheng Kung University, Taiwan, and the National Center for Theoretical Sciences (NCTS), Taiwan. J.L. acknowledges support from the National Science Foundation of China (Grant No.11674375, 11634015) and the National Basic Research Program of China (Grant No. 2015CB921300, 2017YFA0302901). M.Z.H. acknowledges support from Lawrence Berkeley National Laboratory and the Miller Institute of Basic Research in Science at the University of California, Berkeley in the form of a Visiting Miller Professorship.

**Materials and Methods**

Single crystal samples were cleaved mechanically *in situ* at 77K in ultra-high vacuum conditions. An iron wire of 99.995% purity was degassed in vacuum and heated to an appropriate temperature and iron was deposited on the sample for a controlled amount of time to obtain desired Fe adatom surface concentration. The samples were then immediately inserted into the STM head, already at He4 base temperature of 4.2K. Tunneling conductance spectra were obtained with chemically etched Pt/Ir tips using standard lock-in amplifier techniques with $V_{RMS}$ of 0.05meV to 0.1meV and a lock-in frequency of 933Hz. The topographic images were taken with tunneling junction set up: V=-50mV, I=200pA. The tunneling conductance spectra were taken with V=-10mV, I=500pA.

To obtain the deconvoluted spectra in Fig. 3(h) we first simulate the curve using the Dynes function [1] and the known superconducting gap structure of $PbTaSe_2$ measured with a normal tip [2] to obtain the gap function of the superconducting tip, modelled by a standard BCS form of a gap size $\Delta_T$ and a thermal broadening from the Fermi-Dirac distribution function at 0.4K. As $\Delta_T$=0.7meV, in between the gap size of bulk Pb and $PbTaSe_2$, it is possible there are some alloying effects with Ta or Se atoms, resulting in the measured $\Delta_T$. We then convolute this tip gap function with a Lorentzian peak at zero energy plus a broadened Dynes gap function of the $PbTaSe_2$ and iteratively fit the parameters until it matches the measured data at Fe impurity within a standard deviation $\chi^2 < 0.01$.